\documentclass[onecolumn,showpacs,floatfix,10pt,nofootinbib]{revtex4}
\pdfoutput=1
\usepackage{epsfig}\usepackage{float}
\RequirePackage{amssymb,amsmath}
\usepackage{makeidx}
\usepackage[latin1]{inputenc}
\usepackage{graphicx}
\usepackage{indentfirst}
\usepackage{color}
\begin{document}

\title{{\Large\texttt The analysis of cross-sections of proton and deuteron induced reactions on
tin isotopes at the beam energy of 3.65 GeV/nucleon.}}
\author{\firstname{A.~R.}~\surname{Balabekyan}}

\email{balabekyan@ysu.am}
\affiliation{%
Yerevan State University, A.Manoogian 1, Yerevan, Armenia
}%
\author{\firstname{N.~A.}~\surname{Demekhina}}
\email{dem_nina@yahoo.com}
\affiliation{%
Yerevan Physics Institute, Alikhanyan st.Yerevan, Armenia
}%

\author{\firstname{V.~M.}~\surname{Zhamkochyan}}
\email{vzh@mail.yerphi.am} \affiliation{Yerevan State University,
A.Manoogian 1, Yerevan, Armenia}

\author{\firstname{G.~S.}~\surname{Karapetyan}}
\email{ayvgay@ysu.am}
\affiliation{Instituto de Fisica, Universidade de Sao Paulo \\
Rua do Matao, Travessa R 187, 05508-900 S~ao Paulo, SP, Brazil
}%


\begin{abstract}

In the given paper the total inelastic cross-sections of the
reactions of protons and deuterons on nuclear targets of enriched
tin isotopes were compared. The factorization of cross-sections of
reactions was discussed. Furthermore, the comparison of theoretical
estimations on total inelastic cross-sections with corresponding
experimental ones was made.

\end{abstract}

\maketitle

\section{Introduction}

In recent years, there has been a considerable interest in the study
of high energy proton-nucleus and nucleus-nucleus collisions. It has
been suggested that such interactions, in addition to extending our
knowledge of nuclear reactions, might also shed some light on the
fundamental questions of nuclear physics.

The interest in the study of spallation reactions has been revived
primarily by the advent of new projects in nuclear technology, in
particular (existence) by investigations on the accelerator-driven
subcritical nuclear power reactors. It is becoming increasingly
evident that, besides having a rigid theoretical background, the
scientists in charge of these projects need to have comprehensive
information concerning the properties of the spallation reactions.
It is assumed that of special interest might be to know to know how
these properties vary with the parameters of the collision.
Currently there are a number of theoretical descriptions of
interaction of high-energy particles with nuclei. The standard
INC+evaporation model describes well the spallation reactions in the
wide range of energy of projectile \cite{brash}. Likewise, the
Glauber model \cite{glauber} also provides quite a sufficient
description of the inelastic scattering of high energy particles on
atomic nuclei. The goal of this paper is to study the reaction
mechanism by comparing the total inelastic  cross-sections of
reactions induced by protons and deuterons on enriched tin isotopes.

\section{Experimental setup and Discussion}

The experiments were conducted on the  Nuclotron of JINR
\cite{bal1,bal2}. The targets of enriched tin isotopes ($^{112}Sn$,
$^{118}Sn$, $^{120}Sn$ and $^{124}Sn$) were irradiated with the
extracted  deuterons beam accelerator. For the study of
cross-sections the method of induced activity was used.
Characteristic gamma spectra were measured on high-purity germanium
detectors. The energy of  protons and deuterons beams was 3.65
GeV/nucleon. On average, 90 residual nuclei were obtained from each
target. The systematization was conducted using 10 parametrical
formula \cite{porile}.
\begin{equation}
\sigma(Z,A)=exp(a1+a2\cdot A+a3\cdot A^2+a4\cdot A^3+(a5+a6\cdot
A+a7\cdot A^2)|Z_p-Z|^{a8})
\end{equation}

\noindent where $Z_p=a9\cdot A+a10\cdot A^2$ is the most probable
charge at the given mass number $A$. The parameters $a1,a2,a3,a4$
determine the shape of the mass curve, the parameters $a5,a6,a7$
determine the width of the charge dispersion curve and the
parameters $a9,a10$ determine the position of the peak of the
distribution of isobaric yields. Fitting was carried out by the
least square method by the program ÇFUMILIÈ. Total inelastic cross-sections for targets are defined as the sum of the isobaric cross-sections $\sigma(Z,A)$.

The estimated total inelastic cross-sections with the theoretical
calculations are presented in Table. Figures 1 and 2 show the
experimental total inelastic cross-section as well as the
theoretical calculations on the mass number of targets. Theoretical
estimates are made via the standard cascade-evaporation model
\cite{brash}, the Glauber model \cite{glauber} and an empirical
expression traditionally used for the data interpretation based on
geometrical predictions  \cite{heck}:

\begin{equation}
\sigma_{tot}=\pi r_0^2 (A_{p}^{1/3}+A_{T}^{1/3}-b)^2
\end{equation}

\noindent where b is overlap parameter, $A_p, A_T$  are the mass
numbers of projectile and target. In Glauber approach \cite{glauber}
the inelastic cross-section of hadron-nucleus interaction is given
by the expression:
\begin{equation}
\sigma_{hA}^{inel}=
\int{d^2b[1-exp(-\sigma_{hN}^{tot}}\int\limits_{-\infty}^{\infty}
\rho (\vec{b},z)dz)]
\end{equation}

where $\rho(\vec{b},z)$ is one particle nuclear density,
$\sigma_{hN}^{tot}$ cross-section of the interaction of incident
hadron $h$ with nucleus nucleon $N$. In calculations the fermi
parametrization was used for $\rho(\vec{r})$ with parameters from
\cite{mur}.

As is seen from the Table I and Figures 1 and 2, for proton-nuclear reactions
theoretical calculations more realistically describe the
experimental results. In the case of the deuterons-induced
reactions, most of experimental points lie above the theoretical
curves, although within the errors they describe the experimental
results. This can be explained by the fact that in the theoretical
calculations some of the effects, related to the nucleon composition
of the projectile, may be not considered.

The concepts of factorization have been developed for the
interpretation of the reaction induced by high energy particles and
nuclei. For the beam energies greater than 2 AGeV single particle
inclusive spectra of target fragments were predicted to depend on
the nature of projectile via the total reactions of cross-section.
The single particle inclusive reaction can be written as $P+T
\rightarrow F+X$, where $P$ and $T$ correspond to the projectile and
target, and $F$ and $X$ to the fragment, produced during the
reaction, and anything else.  If the hypothesis of factorization
exists the cross-section for the product of target fragmentation $F$
can be factorized to  $\sigma_{T,P}^F=\sigma_T^F \gamma_p$,  where
$\gamma_p$ is dependent only on the projectile. If we have proton-
and deuteron-induced reaction we can write
$$\sigma^F(d+Sn)/\sigma^F(p+Sn)=\gamma_d/\gamma_p=R,$$
where $R$ is a relative projectile factor. If the factorization
hypothesis is valid, this factor should be equal to that of the
total reaction cross-sections.

Figure 3 shows the ratio (R) of cross-sections of residual nuclei in
deuteron-nucleus reactions to the proton-nucleus reactions for all
targets. As can be seen for targets similar mass to the natural
composition of the isotope ($^{118}Sn, ^{120}Sn$) these relations on
average are in agreement with published data (R=1,6) \cite{kozma}.
For targets $^{112}Sn$ and $^{124}Sn$ cross-section the ratio are
different, in particular, for $^{112}Sn$, this ratio is 1.3, and for
$^{124}Sn$, it is 1.9. One should expect that the ratio of the cross-section of the residual nuclei for deuteron- and proton-nucleus
reactions is sensitive to the isotope composition of the target-nucleus. In the Table I, together with the total inelastic cross-section are the radii of the projectile and the target nucleus, as
well as the effective impact parameters of interaction computed
according to the formula (2). Since $1/2(R_p+R_T)<b$ for all the
targets we can conclude that the collision is peripheral. This fact
can be explained as follows. The  difference in the interaction of
deuterons  and  protons should  be revealed  in the transferred
excitation energy. This process is dependent on the collisions
conditions  and the impact parameters values. In peripheral
collisions at large impact parameter the probability of the whole
deuteron interaction is very low. Taking into account  the low
coupling energy  in deuteron it can be presented that in this case
only one  participant  nucleon from  deuteron  interacts   with
target. The target spallation , which takes place at relatively low
excitation energies, is more probable in these cases. The cross
sections  with low number emitted nucleons  are the largest and
carry the largest statistical weight. The experimental cross
sections  do not change in energy range  about several GeV, at least
for proton-induced reactions. In other words spallation cross-sections at high energy are saturated in regime of limiting
fragmentation. Taking into account the above mentioned discussion it
is clear, that cross-section ratio in the target spallation range  
($A\geq 60$ amu) will be very similar for proton- and deuteron-induced reactions at the energies of a few GeV. On average 
$\sigma_d(A,Z)/\sigma_p(A,Z)=1.0\pm0.1$
 for all targets. At higher
excitation energies the multifragmentation phenomenon appears, these
cases are related to the more central interactions of the incident
projectiles and target. The cross sections of these reactions are
small  and statistical uncertainties are essentially increased  (as
can be seen from fig. 3). The statistical weights of these
collisions are low. For this reason they do not affect on the
average values of the calculated impact parameter. But these cases
can present the deuteron deposit in the total interaction process.
The cross-section ratio in the case of deuteron to proton for these collisions
essentially exceeds the value near one in spallation range
$\sigma_d(A,Z)/\sigma_p(A,Z)=2.0\pm0.2$ for all targets.

\section{Conclusion}
The total inelastic cross sections for proton-nuclear and
deuteron-nuclear  reactions   with theoretical calculations were
compared. The data received permits to conclude that for
proton-nuclear reactions theoretical calculations
 more realistically describe the experimental results.
The impact parameter of interaction of protons and deuterons with
tin targets were obtained. According to the values of these
parameters we conclude that the collision is peripheral.

\section*{Acknowledgment} G. Karapetyan is grateful to Funda\c c\~ao
de Amparo \`a Pesquisa do Estado de S\~ao Paulo (FAPESP)
2011/00314-0. and to the International Centre for Theoretical
Physics (ICTP) under the Associate Grant Scheme.

\newpage
\vskip 4cm
\begin{figure}[h!]
\includegraphics[scale=0.16]{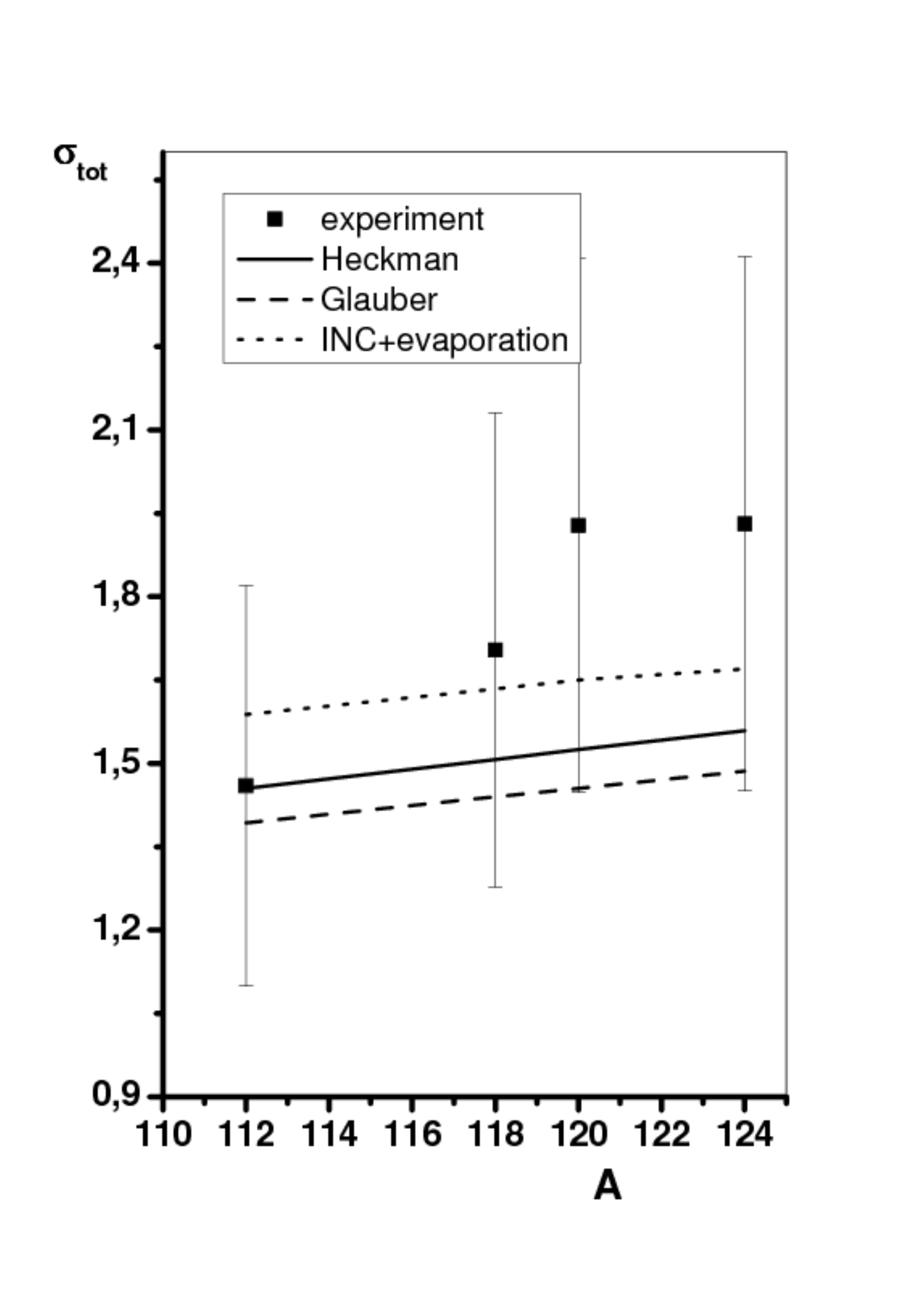}
\vskip -3cm \caption{\small Total inelastic cross sections of
deuteron-nucleus interactions}
\end{figure}

\vskip 4cm
\begin{figure}[h!]
\includegraphics[scale=0.16]{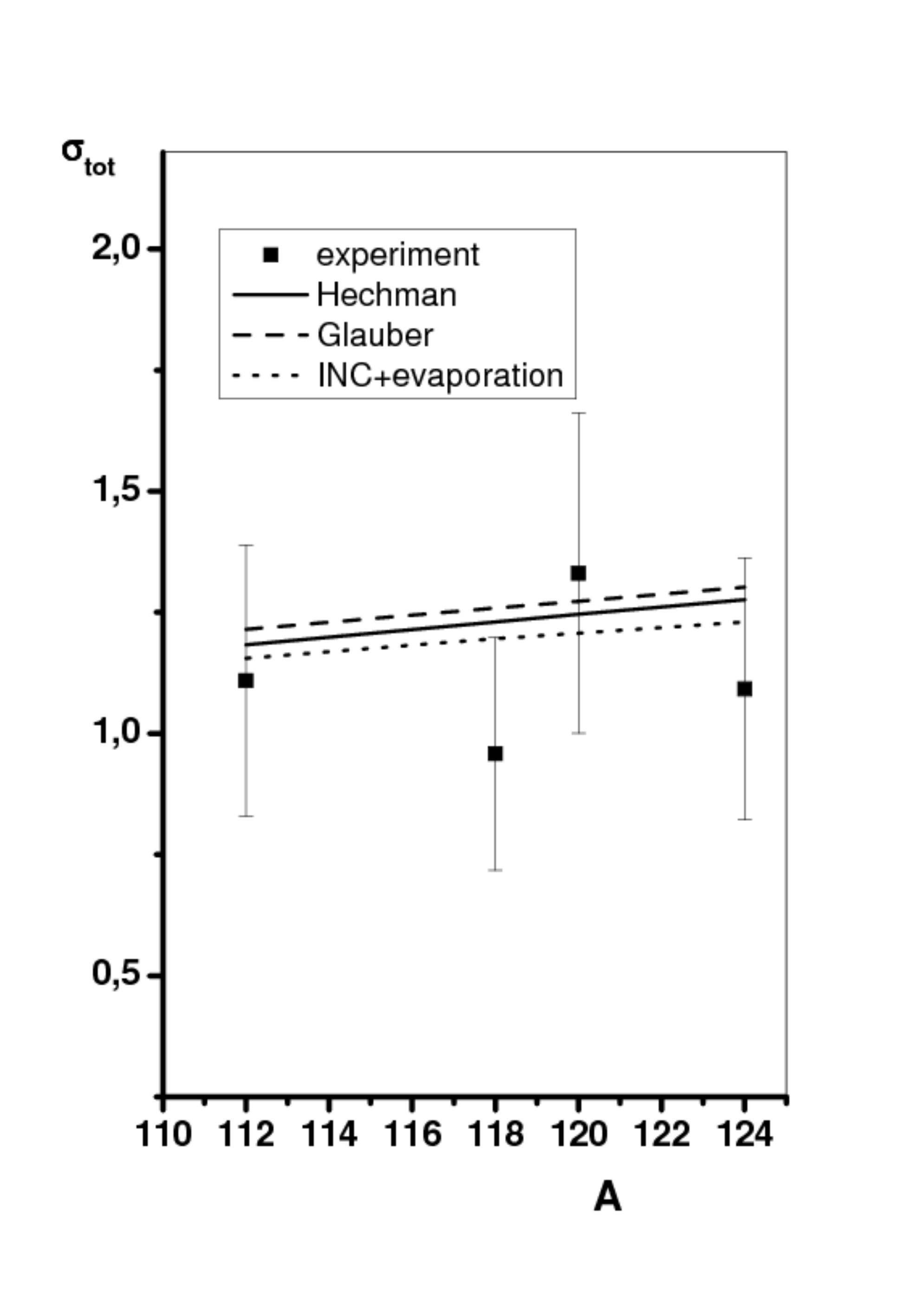}
\vskip -3cm \caption{\small Total inelastic cross sections of
proton-nucleus interactions}
\end{figure}

\vspace*{25mm}
 \begin{figure}[h]
 \hspace*{-30mm}
 \includegraphics[scale=0.8]{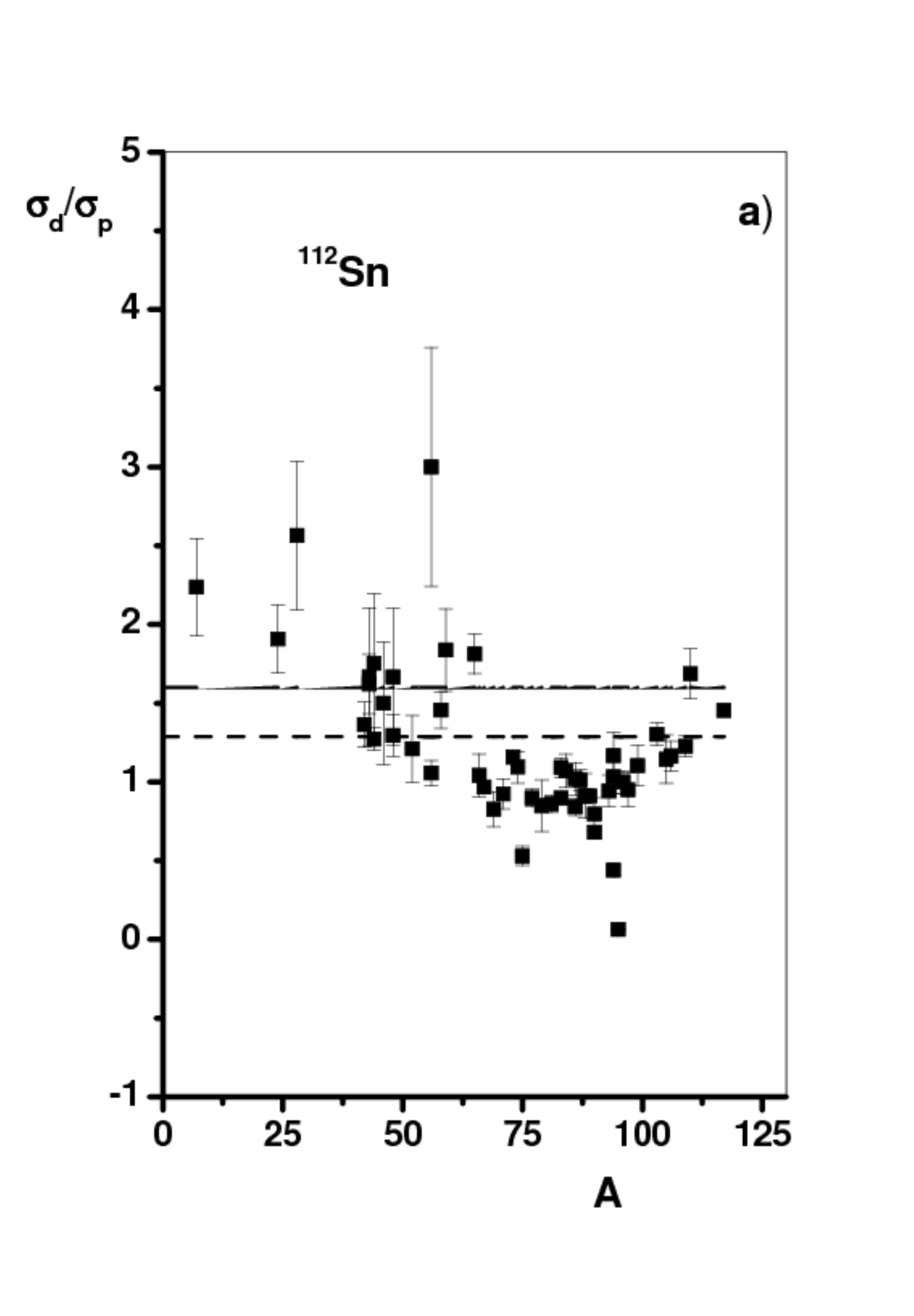}
 \hspace*{-15mm}
 \includegraphics[scale=0.8]{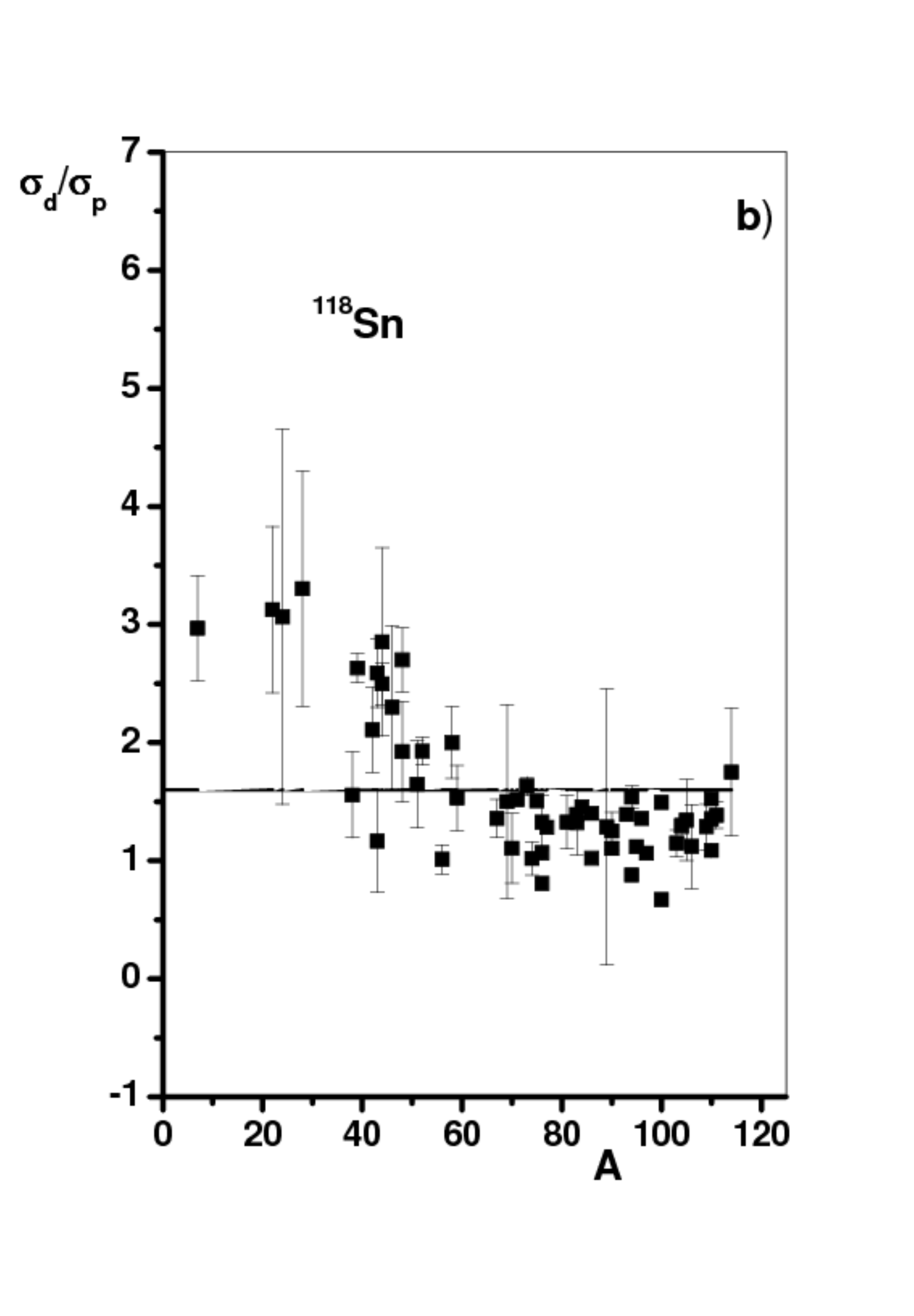}
 \hspace*{-30mm}
 \includegraphics[scale=0.8]{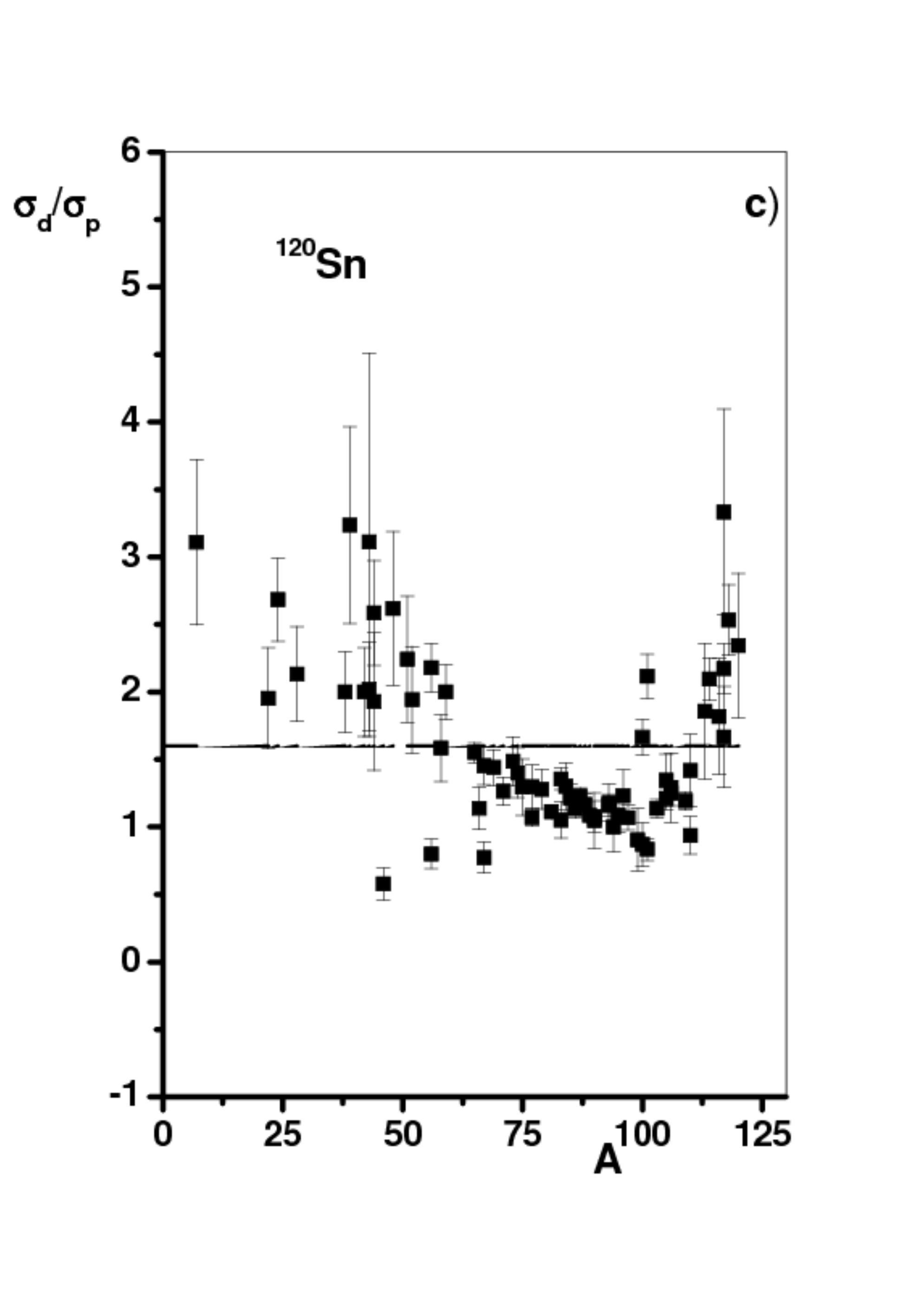}
 \hspace*{-15mm}
 \includegraphics[scale=0.8]{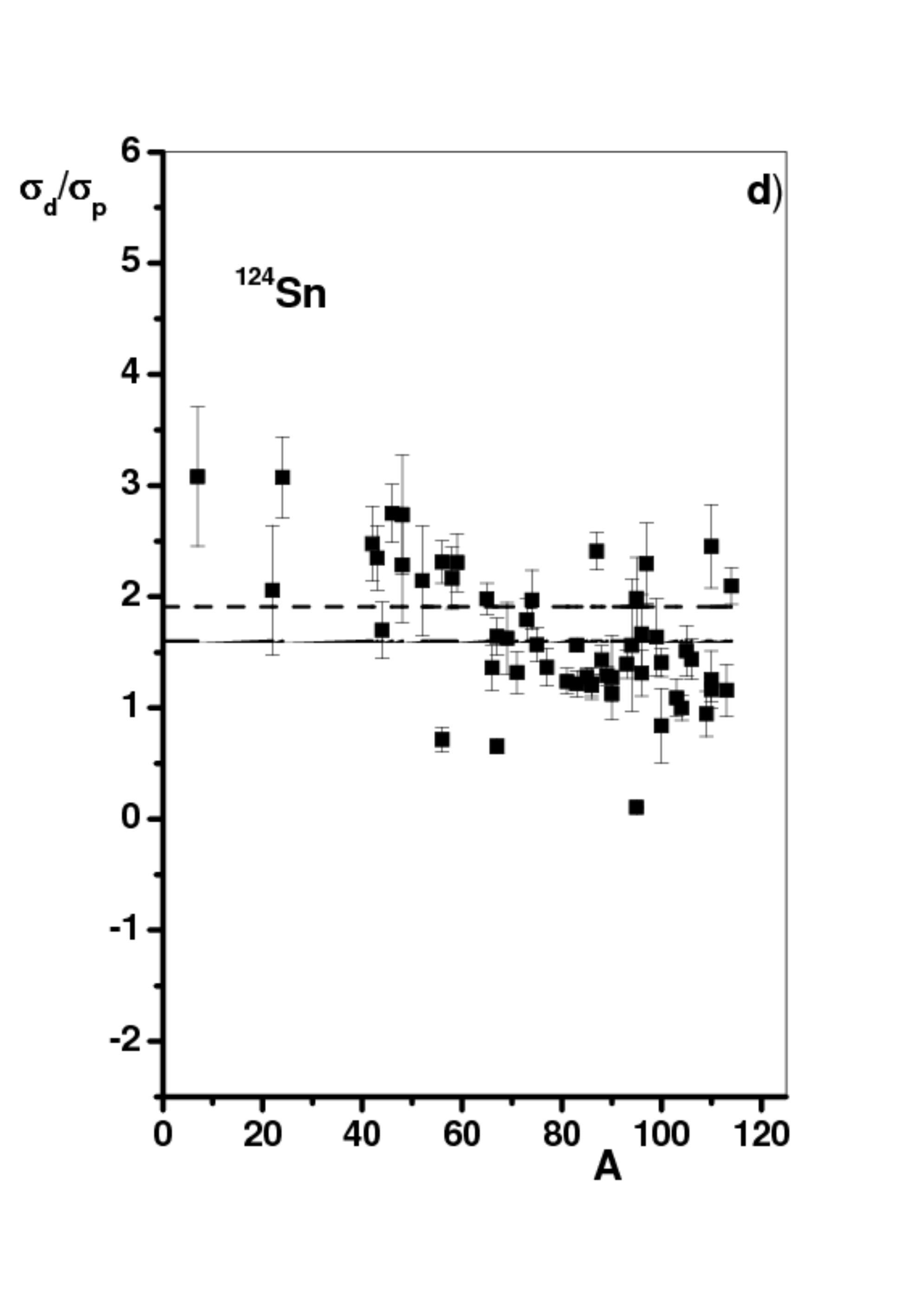}
 \vspace*{-35mm}
 \caption{The ratio of deuteron induced to proton induces reactions cross sections}
 \end{figure}

\newpage
\begin{table}[!p]
Table. The total inelastic cross-sections of deuterons and protons
induced reactions and the overlap parameters. \vskip 5mm
\begin{tabular}{|c|c|c|c|c|c|c|c|} \hline
reactions& exp. & Heck. &   Glauber   & INC &  R$_p$ &
R$_T$& parameter b \\
& bn& bn & bn& bn & fm & fm & fm \\ \hline $p+^{112}Sn$ &1.109$\pm$
0,27 & 1.183& 1.215 &  1.155 &  1.37  &  6.604 &  6.43 \\ \hline
$p+^{118}Sn$ &0,951$\pm$ 0,24 & 1.23& 1.259& 1.196 &  1.37 &  6.72
&6.45 \\ \hline $p+^{120}Sn$ &1.331$\pm$ 0,33& 1.246& 1.273& 1.207&
1.37 &6.757& 7.35 \\ \hline $p+^{124}Sn$ & 1.092 $\pm$ 0,27& 1.276 &
1.302&  1.23& 1.37& 6.83& 6.62 \\ \hline $d+^{112}Sn$ & 1.46$\pm$
0,36 &1.455& 1.393& 1.588 &1.73& 6.604 &7.84 \\ \hline $d+^{118}Sn$
& 1.704$\pm$ 0,43& 1.507& 1.44 &1.634& 1.73& 6.72& 8.21\\ \hline
$d+^{120}Sn$ & 1.928$\pm$ 0,48& 1.525& 1.455& 1.65 &1.73& 6.757& 8.4
\\ \hline
$d+^{124}Sn$ &1.931$\pm$ 0,48&  1.559& 1.486& 1.67& 1.73& 6.83& 8.46
\\ \hline
\end{tabular}
\end{table}

\end{document}